\documentclass[10pt]{amsart}

\usepackage{tikz}
\usepackage{amssymb,amsmath,latexsym,graphicx}
\usepackage{charter,eucal}
\usepackage{amssymb}
\usepackage{amscd}

\newtheorem{theorem}{Theorem }[section]
\newtheorem{lemma}[theorem]{Lemma}
\newtheorem{remark}[theorem]{Remark}
\newtheorem{corollary}[theorem]{Corollary}
\newtheorem{proposition}[theorem]{Proposition}
\newtheorem{question}[theorem]{\textsc{Question}}
\newtheorem{conjecture}[theorem]{\textsc{Conjecture}}
\newtheorem{definition}[theorem]{\textsc{Definition}}

\def\1{\mathrel{\mathbf{1}}}

\newcommand{\bI}{\mathrm{id}}

\newcommand{\eop}{\hspace*{\fill}$\blacksquare$}

\newcommand{\btt}{\begin{ttheorem}}
\newcommand{\ett}{\end{ttheorem}}

\newcommand{\bt}{\begin{theorem}}
\newcommand{\et}{\end{theorem}}
\newcommand{\bcc}{\begin{conjecture}}
\newcommand{\ecc}{\end{conjecture}}
\newcommand{\bc}{\begin{corollary}}
\newcommand{\bl}{\begin{lemma}}
\newcommand{\ec}{\end{corollary}}
\newcommand{\el}{\end{lemma}}
\newcommand{\bq}{\begin{question}}
\newcommand{\eq}{\end{question}}
\newcommand{\bp}{\begin{proposition}}
\newcommand{\ep}{\end{proposition}}
\newcommand{\br}{\begin{remark}}
\newcommand{\er}{\end{remark}}
\newcommand{\bd}{\begin{definition}}
\newcommand{\ed}{\end{definition}}

\newcommand{\mW}{\ensuremath{\mathcal{W}}}

\newcommand{\PG}{\ensuremath{\mathbf{PG}}}

\newcommand{\mC}{\ensuremath{\mathcal{C}}}

\newcommand{\hP}{\ensuremath{\mathbf{P}}}
\newcommand{\mU}{\ensuremath{\mathcal{U}}}

\newcommand{\mB}{\mathcal{B}}

\author{Koen  Thas}

\address{Department of Mathematics,
Ghent University,
Krijgslaan 281, S25, B-9000 Ghent, Belgium}

\email{koen.thas@gmail.com}

\title[Mutually unbiased bases]{On the mathematical foundations of mutually unbiased bases}

\date{}

\begin{document}

\maketitle

\begin{abstract}
In order to describe the right setting to handle Zauner's conjecture on mutually unbiased bases (MUBs) (saying that in $\mathbb{C}^d$, a
set of MUBs of the theoretical maximal size $d + 1$ exists only if $d$ is a prime power), we pose
some fundamental questions which naturally arise. Some of these questions have important consequences for the construction theory of (new) sets of maximal MUBs.
\end{abstract}

{\bf PACs numbers}: 02.10.Hh, 02.40.Dr, 03.67.-a, 03.65.Ta, 03.65.Ud\\

\bigskip
\section{Introduction}


Two orthonormal bases $\mB$ and $\mB'$ of the Hilbert space $\mathbb{C}^\ell$ ($\ell \in \mathbb{N}^{\times}$) are {\em mutually unbiased}
if and only if

\[ \vert \langle \phi \vert \psi \rangle \vert^2 = 1/\ell            \]
\noindent
for all $\vert\phi\rangle \in \mB$ and $\vert\psi\rangle \in \mB'$.
It is a fundamental and very famous conjecture, sometimes referred to as ``Zauner's conjecture'' \cite{Zauner} (although quite probably the conjecture already floated around before 1999),  that the theoretical upper bound 
$\ell + 1$ of a set of mutually unbiased bases (``MUBs'')  can only be reached when $\ell$ is a {\em prime power}. For each such $\ell$,
examples exist | in fact, there is a rich literature in the construction theory of such examples, and even for $\ell = 6$, the first case for which the conjecture is open, many papers exist.

MUBs were introduced by Julian Schwinger in 1960 \cite{Schwinger} under a different name. He noted in \cite{Schwinger} that bases which are mutually unbiased represent measurements that are maximally non-commutative, in the sense that a measurement over one such basis leaves one completely uncertain as to the outcome of a 
measurement over a basis which is mutually unbiased with the first. Later, in \cite{wf} Wootters and Fields introduced the term ``mutually unbiased bases.''

In an attempt to better understand the category of (maximal sets of) MUBs, and more precisely, to find the ``correct setting'' to attack 
Zauner's conjecture, some (rather subtle) questions have popped up very naturally which might be interesting in their own right (both from a 
physical and mathematical point of view). 

 For instance, one of the main tools in the construction theory of maximal sets of MUBs, after \cite{Band}, is the theory of so-called ``maximal commuting operator classes'' (MCCs). From such an MCC (of size $d + 1$), a maximal set of MUBs (of size $d + 1$) can be derived, and from a maximal set of MUBs (of size $d + 1$) one can also make an MCC (of size $d + 1$). We will show that one has to be very careful when constructing ``new'' maximal MUBs
through the theory of MCCs, as {\em nonisomorphic} MCCs could give the {\em same} maximal set of MUBs! 

Several questions on the correspondence between MCCs and MUBs will thus be formulated.

 These (and the aforementioned) questions are the subject of the present letter (which at the same time can be considered as a first installment in a series of papers on Zauner's conjecture).

\bigskip
\section{MUBs and maximal commuting operator classes}
\label{UMUBdef}

Let $\mU$ be a set of $d^2$ mutually orthogonal unitary operators in $\mathbb{C}^d$ using the Hilbert-Schmidt norm: operators $A$ and $B$ are {\em orthogonal} if $\mathrm{tr}(AB^{\dagger}) = 0$. 
Along with the identity operator $\bI$, $\mU$ constitutes a basis for the $\mathbb{C}$-vector space of $(d \times d)$-complex matrices $\mathbf{M}_{d\times d}(\mathbb{C})$.
A standard construction of MUBs outlined in \cite{Band} relies on finding classes of commuting
operators, with each class containing $d - 1$ mutually orthogonal commuting unitary matrices
different from the identity $\mathrm{id}$.\\

A set of subsets $\{\mC_1, \mC_2,\ldots, \mC_{\ell} \vert \mC_j \subset \mU \setminus \{\bI\}\}$ of size 
$\vert \mC_j \vert = d - 1$ constitutes a (partial) partitioning of $\mU \setminus \{\bI\}$ into {\em mutually
disjoint maximal commuting classes} if the subsets $\mC_j$ are such that 
\begin{itemize}
\item[(a)]
the elements of $\mC_j$
commute for all $1 \leq j \leq \ell$ and 
\item[(b)] 
$\mC_j \cap \mC_k = \emptyset$ for all $j \ne k$.
\end{itemize}

If $\ell = d$, we speak of an ``MCC.''


\bl[\cite{Band}]
The common eigenbases of $\ell$ mutually disjoint maximal commuting operator
classes form a set of $\ell$ mutually unbiased bases.
\el

For the rest of this paper, if $\mB$ is a set of MUBs of size $\ell + 1$ in $\mathbb{C}^{\ell}$, we will call $\mB$ a {\em maximal set of MUBs}
or a {\em maximal MUB} for short.

Define the map $\beta$ from the set of all MCCs, denoted $\mathbf{MCC}$, to the set of all sets of maximal MUBs, denoted $\mathbf{MUB}$, as being the map which sends an MCC $\mU$ to 
the set of MUBs $\beta(\mU)$ which arise as common eigenbases as in the previous lemma.

Conversely, consider a set $\mB$ of $d + 1$ MUBs in $\mathbb{C}^d$. Denote its $d + 1$ bases by $\mB_0,\mB_1,\ldots,\mB_d$, and 
for each $i = 0,1,\ldots,d$, let 
\begin{equation}
\mB_i = \{ \langle \psi_1^i \vert, \langle \psi_2^i \vert, \ldots, \langle \psi_d^i \vert \}.
\end{equation}

Following \cite{Band}, define for each $k = 0,1,\ldots,d$ and $j = 1,2,\ldots,d$,
\begin{equation}
U_j^k = \sum_{r = 1}^de^{2\pi ijr/d}\vert \psi^k_r\rangle \langle \psi_r^k \vert.
\end{equation}
Then with $\mU_y$, $y = 0,1,\ldots,d$, defined as $\{ U^y_j\vert j = 1,2,\ldots,d \}$, 
$\{\mU_0, \mU_1, \ldots, \mU_d\}$ is an MCC of size $d + 1$, denoted $\alpha(\mB)$ in this letter. 

So $\alpha$ defines a map between the set of  maximal MUBs and the set of MCCs.

As a corollary of the existence of $\alpha$ and $\beta$, we have that {\em a maximal set of MUBs exist in $\mathbb{C}^d$ if and only if 
an MCC of size $d + 1$ exists}. \\

The first questions we want to pose concerns the nature of the compositions $\alpha \circ \beta: \mathbf{MCC} \longrightarrow \mathbf{MCC}$ and $\beta\circ \alpha: \mathbf{MUB} \longrightarrow \mathbf{MUB}$.

We start with introducing the general Pauli group.

\medskip
\subsection{The general Pauli group}

Let $d$ be a prime.
Let $\{ \vert s \rangle \vert s = 0,1,\ldots,d - 1\}$ be a computational base of $\mathbb{C}^d$.
Define the $d^2$ (generalized) {\em Pauli operators} of $\mathbb{C}^d$ as

\[  (X_d)^a(Z_d)^b,\ \ \ a,b \in \{0,1,\ldots,d - 1\},         \]
\noindent
where $X_d$ and $Z_d$ are defined by the following actions

\[  X_d \vert s\rangle = \vert s + 1\mod{d}\rangle, \ \ \ Z_d\vert s\rangle = \omega^s \vert s\rangle,        \]
\noindent
where {$\omega =$ exp$(2i\pi/d)$}.

\medskip
The set $\mathbb{P}$ of generalized Pauli operators of the $N$-qudit Hilbert space $\mathbb{C}^{d^N}$ is the set $\mathbb{P}$ of $d^{2N}$ distinct tensor products of the form

\[ \sigma_{i_1} \otimes \sigma_{i_2} \otimes \cdots \otimes \sigma_{i_N},             \]
\noindent
where the $\sigma_{i_k}$ run over the set of (generalized) Pauli matrices of $\mathbb{C}^d$. Denote $\mathbb{P}^{\times} = \mathbb{P} \setminus \{\bI\}$.
These operators generate a group under ordinary matrix multiplication, denoted $\hP = \hP_N(d)$ and called the {\em general Pauli group} (or {\em {discrete} Heisenberg-Weyl group}).
 It has order $d^{2N + 1}$.

 In the following proposition, $[.,.]$ denotes the commutator relation in the group $\hP$.

\begin{proposition}[\cite{Appl}]
\label{prop}
\begin{itemize}
\item[{\rm (i)}]
The derived group $\hP' = [\hP,\hP]$ equals the center $Z(\hP)$ of $\hP$.
\item[{\rm (ii)}]
We have $Z(\hP) = \langle \omega\cdot\bI\rangle$, so that $\vert Z(\hP)\vert = d$.
\item[{\rm (iii)}]
$\hP$ is nonabelian of exponent $d$ if $d$ is odd; if $d = 2$, $\hP$ is nonabelian of exponent $4$.
\item[{\rm (iv)}]
We have the following short exact sequence of groups:
\[    1 \mapsto Z(\hP) \mapsto \hP \mapsto V(2N,d) \mapsto 1.            \]
\end{itemize}
\end{proposition}

Observe that $\hP/Z(\hP)$ can be identified with $\mathbb{P}$ ($Z(\hP)$ corresponds to the identity operator).

Denote the natural map $\hP \mapsto V(2N,d)$ by an overbar. Then the commutator

\[ [.,.]: V(2N,d)\times V(2N,d)  \mapsto \langle \omega\cdot\bI\rangle:  (\overline{v_1},\overline{v_2}) \mapsto [ \overline{v_1},\overline{v_2}] = [v_1,v_2] \]
\noindent
defines a {non-degenerate} alternating bilinear form on $V(2N,d)$ (the derived group $\hP'$ is identified with the additive group of $\mathbb{F}_d$), so also on the corresponding projective space $\PG(2N - 1,d)$.

\medskip
\subsection{Symplectic polar spaces and the Pauli group}

Now consider the projective space $\PG(2N - 1,d)$ of dimension $2N -
1$, $N \geq 2$, over the field $\mathbb{F}_d$ with $d$ elements. Let $F$ be a
{non-degenerate} symplectic form of $\PG(2N - 1,d)$. 
For $F$ one can choose the following canonical bilinear form \cite{Hirsch}:

\[ (X_0Y_1 - X_1Y_0) + (X_2Y_3 - X_3Y_2) + \cdots + (X_{2N - 2}Y_{2N - 1} - X_{2N - 1}Y_{2N - 2}).          \]

Then the {\em symplectic polar space} $\mW_{2N -
1}(d)$ consists of the points of $\PG(2N - 1,d)$ together with all
totally isotropic spaces of $F$ \cite{Hirsch}. Here, a {\em totally isotropic subspace} is 
a linear subspace $W$ of $\PG(2N - 1,d)$ that vanishes under $F$ (i.e., the restriction of $F$ to $W$ is trivial).

By the previous subsection, the general Pauli group $\hP_N(d)$ naturally defines a symplectic polar space $\mW_{2N - 1}(d)$. 
There is a natural surjective map
\[ \gamma: \mathbb{P}^\times  \longrightarrow  \mbox{points}\ \mbox{of}\ \mW_{2N - 1}(d) \]
such that  operators $x$ and $y$ commute if and only if 
the  points $\gamma(x)$ and $\gamma(y)$  of $\mW_{2N - 1}(d)$ generate a linear subspace which vanishes under $F$  | see \cite{Appl,KT-recent} for more details.

Now if $S$ is a {\em partition} of $\mW_{2N - 1}(d)$ in totally isotropic subspaces of (maximal) dimension $N - 1$, then $\gamma^{-1}(S)$ defines
an MCC of size $d^N + 1$ (also denoted $\gamma^{-1}(S)$), see \cite{Appl,KT-recent} for details and also Remark \ref{remgam} below, and hence a maximal MUB in $\mathbb{C}^{d^N}$. 
Many such examples exist. For further reference, we will call a partition of the aforementioned type a {\em spread}.

\br{\rm
If $\sigma \in \mathbb{P}^{\times}$, $\langle \sigma \rangle$ is the vector line generated by $\sigma$ in $V(2N,d) \equiv \hP/Z(\hP) \equiv \mathbb{P}$, and this line maps to a point of $\mW_{2N - 1}(d)$.
}
\er

\br[On $\gamma^{-1}(\cdot)$ and scaling]
\label{remgam}
{\rm
Let $S$ be as above, and consider again $\gamma^{-1}(S)$. In \cite{Appl,KT-recent} it is shown that it defines precisely one MCC on the set
of nontrivial Pauli operators $\mathbb{P}^\times = \{ \nu_1,\ldots,\nu_{d^{2N} - 1}\}$ in $\hP$, and this particular MCC is the one we consider. On the other hand, if 
$c_1,\ldots,c_{d^{2N} - 1}$ are arbitrary $d$-th roots of unity in $\mathbb{C}$, then $\gamma^{-1}(S)$ also defines an MCC on 
$\{ c_1\nu_1,\ldots,c_{d^{2N} - 1}\nu_{d^{2N} - 1}\}$ (and this remains valid in general; we will call this process ``scaling''). But this MCC obviously should be isomorphic to the one defined on Pauli operators in any good theory of morphisms for MCCs. 
}
\er

\medskip
\subsection{The maps $\alpha$ and $\beta$}

Take an element $\mB$ in $\mathbf{MUB}$, and consider $\alpha(\mB) = \{ \mU_0,\mU_1,\ldots,\mU_d\}$ (we use the notation of above).
As each element $U^j_1$ with $j = 0,1,\ldots,d$ has $d$ different eigenvalues, it follows that up to scaling the image of $\alpha(\mB)$
under $\beta$ is unique (that is, is $\mB$ again). So $\alpha$ is injective. 

In order to understand the correspondence between maximal MUBs and MCCs, we need to understand the map $\beta$ as well. What first comes to mind is the question whether $\beta$ is injective | i.e., could it happen that two different (nonisomorphic) MCCs give rise to the same maximal MUB
under $\beta$? When attacking Zauner's conjecture from the viewpoint of MCCs, it would be very valuable to have a canonical correspondence
between maximal MUBs and MCCs, but unfortunately, $\beta$ is {\em not} injective: {\em very} structurally different MCCs could map to 
the same maximal MUB. As each MCC generates a group, this means for instance that nonisomorphic groups can carry the same MUB structure.

This has important implications for the construction theory of maximal MUBs: one has to be very careful when constructing ``new'' maximal MUBs
through the theory of MCCs (and the map $\beta$), as nonisomorphic MCCs could give the same MUB!

\br
{\rm
Note that for any maximal MUB $\mB$, $\beta^{-1}(\mB)$ is not empty, since $\alpha(\mB) \in \beta^{-1}(\mB)$. So $\beta$ is surjective.}
\er

Before proceeding, we need to express what ``isomorphic MCCs'' means. (In \cite{KT-recent} this notion was already discussed in the special
case of MCCs consisting of Pauli operators; there, a finer definition can be given than the one we propose here in the general context.)

For any MCC $\widetilde{\mU}$ of size $d + 1$,
define $A(\widetilde{\mU}) \leq \mathbf{U}_d(\mathbb{C})$ to be the group generated by the elements of  $\cup_{i = 0}^d\widetilde{\mU}_i$.

So let $\mU$ and $\mU'$ be MCCs, and let $\Omega$, respectively $\Omega'$, be the set of operators of $\mU$, respectively
$\mU'$. 
Then we call $\mU$ and $\mU'$ {\em isomorphic} if modulo scaling there exists an isomorphism $\varrho: A(\mU) \mapsto A(\mU')$ which maps $\Omega$ to $\Omega'$. (By ``modulo scaling'' we mean that one first is allow to re-scale $\Omega$.)\footnote{Probably other (better) definitions exist, but it is in any case compatible with the one of \cite{KT-recent} in the special case of Pauli operators.}

\bq
Consider a maximal MUB $\mB$. List invariants of the elements of $\beta^{-1}(\mB)$. 
\eq

Now let $d$ be a prime, $N > 1$ a positive integer, and let $S$ be a spread of $\mW_{2N - 1}(d)$ (they always exist); then we have seen that $\gamma^{-1}(S)$ is an MCC of size $d^N + 1$, so $\beta(\gamma^{-1}(S)) =: \mB_S$ is a maximal MUB of order $d^N  + 1$. Now $\gamma^{-1}(S)$ and 
$\alpha(\mB_S)$ both are elements of $\beta^{-1}(\mB)$, but they cannot be isomorphic for various reasons. One being that as $\gamma^{-1}(S)$ is 
a subset of the general Pauli group $\hP_N(d)$, all its elements are trivial or have order $d$. While obviously $\alpha(\mB_S)$ contains
elements of order $d^N$ (as each $\mU^j \cup \{\bI\}$ is a cyclic group of order $d^N$).

So the list of all possible orders of the operators associated to an element of $\beta^{-1}(\mB)$ is {\em not} an invariant! Moreover,
$A(\gamma^{-1}(S))$, the Pauli group, is a $d$-group of exponent $d$, while $A(\alpha(\mB_S))$ is a $d$-group of exponent $d^N$, so the associated (isomorphism classes of) groups aren't invariants as well ...

This indicates that the sets $\beta^{-1}(\mB)$ behave rather mysteriously.

\bigskip
\section{A theory of heights?}

In order to work with an induction hypothesis, it could be valuable to introduce a notion of ``height'' for any MCC, which measures how far the generated
group is from an abelian group. Ideally, the heights would be nonzero integers, and height $1$ would be the case where the group {\em is} abelian.

So let $\mU =  \{ \mU_0,\mU_1,\ldots,\mU_d\}$ be an MCC of size $d + 1$. The {\em height} of $\mU$ is the value
\begin{equation}
\rho(\mU) := \frac{\vert A(\mU) \vert}{d^2} \in \mathbb{Q}_{\geq 1}. 
\end{equation}

At this point, I have no idea whether a height is always contained in $\mathbb{N} \cup \{\infty\}$, or even $\mathbb{N}$, and these are the first questions to be handled. Each of the questions below comes with a more subtle twin, motivated by the previous section.

\bq
Let $\mU$ be an MCC of size $d + 1$.
\begin{itemize}
\item[{\rm (a)}]
 Is $\rho(\mU)$ always finite? (That is, is $A(\mU)$ always a finite group?)
 \item[{\rm (b)}]
Is $\rho(\widetilde{\mU})$ finite for some element $\widetilde{\mU} \in \beta^{-1}(\beta(\mU))$? (In other words,
given $\mU$, is there some other MCC $\widetilde{\mU}$ of finite height which gives rise to the same maximal MUB?) 
\end{itemize}
\eq

A positive answer on either questions would reduce the problem to one in finite (unitary) group theory.

\bq
Let $\mU$ be an MCC of size $d + 1$.
\begin{itemize}
\item[{\rm (a)}]
 Is $\rho(\mU)$ always a positive integer, or $\infty$? 
 \item[{\rm (b)}]
 If not all elements of $\beta^{-1}(\beta(\mU))$ have infinite weight,
is $\rho(\widetilde{\mU})$ a positive integer for some element $\widetilde{\mU} \in \beta^{-1}(\beta(\mU))$?  
\end{itemize}
\eq

MUBs which are associated to an MCC of height $1$ can be easily handled; we will do this in the following section.

Taken that $\rho(\mU) \ne \infty$, one way to study these questions could be to consider the subgroups $\langle \mU_i,\mU_j \rangle$ ($i \ne j$)
and try to find out how the commutator $[\mU_i,\mU_j]$ looks like (so that one can estimate the order of $\langle \mU_i,\mU_j \rangle$). It seems that even in special cases this becomes a hard task. This motivates us to consider the nilpotence class of the groups generated by the MCCs associated to one given maximal MUB.

 \bigskip
 \section{MUBs of class $2$}
 
I say that an MCC $\mU$ has {\em class} $m \in \mathbb{N} \cup \{ \infty\}$ if $A(\mU)$ has nilpotence class $m$. The class is denoted by 
$\mathrm{cl}(\mU)$. (Class $\infty$ means that $A(\mU)$ is not nilpotent.)
 
A maximal MUB $\mB$ has {\em class} 
$n \in \mathbb{N} \cup \{ \infty\}$ if $n = \mathrm{min}\{ \mathrm{cl}(\mU) \vert \mU \in \beta^{-1}(\mB) \}$.

If $\mB$ has class $1$, this means that some $\mU \in \beta^{-1}(\mB)$ has class $1$, so that $A(\mU)$ is abelian. 
On the other hand, in \cite{Band} it is remarked that if $O \ne \mathrm{id}$ is an operator in an MCC, $O$ only commutes with 
the elements of the unique commuting class to which it belongs. So $A(\mU)$ cannot be abelian.

\bp
\label{obs}
A maximal MUB always has class at least $2$.\eop 
\ep

 
\bc
If $\mB$ is a maximal MUB of size $d + 1$ and $\mU \in \beta^{-1}(\mB)$, then $\mU$ cannot have height $1$.
\ec
{\em Proof}.\quad
Suppose by way of contradiction that $\mU$ has height $1$; then $\vert A(\mU) \vert = d^2$. So $A(\mU)^{\times} = \cup_{i = 0}^d\mU^i$.
By the remark preceding Proposition \ref{obs}, each commuting class $\mU^j$ generates a subgroup of $A(\mU)$ which must be 
equal to $\mU^j \cup \{\mathrm{id}\}$. So in this way we get $d + 1$ subgroups of $A(\mU)$ of size $d$, two by two intersecting only in $\{ \bI\}$.
It follows by \cite[Theorem 7.6]{HP} that $A(\mU)$ must be abelian, contradiction. \eop \\

Any MCC $\mU$ of order $d^N + 1$ which exists solely of elements from the general Pauli group $\hP_N(d)$ of order $d^{2N + 1}$ (that is, which 
arises from the symplectic polar space), has the property that $A(\mU) = \hP_N(d)$, and as we have seen the general Pauli group has nilpotence class $2$. So class $2$ MUBs {\em do} exist. 

\bq
Can Zauner's conjecture be verified for MUBs of class $2$? 
\eq

Working with general MCCs of class $2$, this already appears to be difficult. I have not verified it, but all known classes of examples
of maximal MUBs should have very low class | in any case, class $2$ maximal MUBs are obviously fundamental to understand.
So we propose the following important special case.

\bq
Let $\mB$ be a maximal MUB of size $d + 1$, and suppose that $\alpha(\mB)$ has class $2$. Is $d$ a prime power?
\eq

 In a forthcoming paper \cite{prep}, we will settle this question in great detail, and classify all such MUBs in a precise way.
 
 I am not sure the following question is in reach.
 
\bq
Let $\mB$ be a maximal MUB of size $d + 1$, and suppose that $\alpha(\mB)$ has class $m < \infty$. Is $d$ a prime power?
\eq

\bigskip
\section{Pauli group in composite dimension}

The considerations of the previous sections aim mostly at eventually finding a positive answer to Zauner's conjecture.
So what about a possible {\em negative} answer | that is, {\em counter examples}? 

One way of trying to construct counter examples | or in any case, to find better lower bounds for the maximal number of 
MUBs in certain composite dimensions | could be a similar construction process such as in \cite{Appl,KT-recent} on the {\em Pauli group in composite dimension}. There have been some allusions on the latter object (in special cases) in \cite{PB}, and a relevant discussion on related abstract groups can be found in \cite{HOS}.

The definition of the ``composite Pauli group'' comes quite naturally, but the question is whether the associated ``symplectic geometry'' can 
be used to construct (not necessarily maximal) sets of MUBs in composite dimension. This geometry would be an amalgamation of smaller 
proper symplectic geometries over the prime fields of which the characteristic occurs in the prime power decomposition of the dimension of the Hilbert space.

I hope to come back to this aspect in a future paper ... \\


\end{document}